\documentclass[%
 reprint,
superscriptaddress,
 amsmath,amssymb,
 aps,
]{revtex4-2}

\usepackage{graphicx}
\usepackage{dcolumn}
\usepackage{bm}
\usepackage[T1]{fontenc}
\usepackage{float}
\usepackage{caption}
\usepackage{subcaption}
\usepackage{afterpage}
\usepackage{xcolor}

\DeclareUnicodeCharacter{2212}{-}
\begin{document}

\newcommand{\mycomment}[1]{}

\preprint{APS/123-QED}

\title{A Ratio-Preserving Approach to Cosmological Concordance} 

\author{Kylar Greene}
 \email{kygreene@unm.edu}
\affiliation{Department of Physics and Astronomy, University of New Mexico \\
Albuquerque, New Mexico 87131}
\affiliation{Theoretical Physics Department, Fermi National Accelerator Laboratory \\
Batavia, Illinois 60510}
\author{Francis-Yan Cyr-Racine}
 \email{fycr@unm.edu}
\affiliation{Department of Physics and Astronomy, University of New Mexico \\
Albuquerque, New Mexico 87131}

\date{\today}

\begin{abstract}

Cosmological observables are particularly sensitive to key ratios of energy densities and rates, both today and at earlier epochs of the Universe. 
Well-known examples include the photon-to-baryon and the matter-to-radiation ratios. 
Equally important, though less publicized, are the ratios of pressure-supported to pressureless matter and the Thomson scattering rate to the Hubble rate around recombination, both of which observations tightly constrain.
Preserving these key ratios in theories beyond the $\Lambda$ Cold-Dark-Matter ($\Lambda$CDM) model ensures broad concordance with a large swath of datasets when addressing cosmological tensions.  
We demonstrate that a mirror dark sector, reflecting a partial $\mathbb{Z}_2$ symmetry with the Standard Model, in conjunction with percent level changes to the visible fine-structure constant and electron mass which represent a \textit{phenomenological} change to the Thomson scattering rate, maintains essential cosmological ratios. 
Incorporating this ratio preserving approach into a cosmological framework significantly improves agreement to observational data ($\Delta\chi^2=-35.72$) and completely eliminates the Hubble tension with a cosmologically inferred $H_0 = 73.80 \pm 1.02$ km/s/Mpc when including the S$H_0$ES calibration in our analysis. 
While our approach is certainly nonminimal, it emphasizes the importance of keeping key ratios constant when exploring models beyond $\Lambda$CDM.

\end{abstract}

\maketitle

\section{\label{sec:intro}Introduction}

The success of the $\Lambda$ cold-dark-matter ($\Lambda$CDM) model at explaining cosmic microwave background (CMB) observations is undeniable, earmarked by its simplicity - six degrees of freedom to explain the Universe \cite{Planck:2018vyg, ACT:2020gnv, SPT-3G:2022hvq}.
Despite its success, $\Lambda$CDM encounters notable discrepancies in reconciling distant cosmic observations with local measurements of the expansion rate today, $H_0$ \cite{Riess:2021jrx,Freedman:2021ahq,Scolnic:2023mrv,Uddin:2023iob}.
The systematic discrepancies between local and distant measurements have been meticulously examined, yet no resolution has been found \cite{Efstathiou:2019mdh,Greene:2021shv,Luongo:2021nqh,Krishnan:2021jmh,Krishnan:2021dyb,Camarena:2022iae,Camarena:2023rsd}. 

Consequently, we investigate beyond the $\Lambda$CDM paradigm to produce concordant cosmologies, drawing on several observational clues as our guide.
The initial hint originates from $\Lambda$CDM directly. 
The model tightly constrains several critical ratios in the early Universe: the baryon-photon ratio, the matter-radiation ratio, the fluid-freestreaming radiation ratio, and the balance between pressure-supported and pressureless matter. 
$\Lambda$CDM's effectiveness at predicting CMB features hinges on its precise predictions of these ratios while maintaining the minimal number of degrees of freedom possible.
The second clue emerges from the significant similarities between the baryon density and the estimated dark matter density, suggesting a potential link in the mechanisms which populate the two sectors.
Given the visible sector's dynamic and complex nature, it is reasonable that if there is a connection between the population of both sectors, the dark sector might be more intricate than just a single particle species.
However, this notion goes against one of $\Lambda$CDM's foundational assumptions.
It assumes that the majority of the Universe's matter consists of a single type of cold, non-interacting, and collisionless dark matter, termed cold dark matter (CDM), detectable solely through its gravitational effects.
Despite its elusive nature and absence from the Standard Model, it is evident that a significant portion of the dark sector must conform to CDM's phenomenological criteria. 
However, the question arises: Must the \textit{entire} dark sector adhere to the CDM paradigm?
Consequently, guided by these clues, our investigation suggests the necessity for a subdominant addition to the dark sector that: maintains the ratios predicted by $\Lambda$CDM and has more dynamic degrees of freedom than CDM.
Separately, we require that the introduction of new physics becomes cosmologically relevant after Big Bang Nucleosynthesis to not interfere with light elemental abundances such as in Refs.~\cite{Berlin:2017ftj,Berlin:2019pbq} where the dark sector thermalizes with the visible sector after BBN.

These clues converge toward a singular solution: a mirror sector inspired by $\mathbb{Z}_2$ symmetries \cite{Arkani-Hamed:2016rle,Chacko:2016hvu,Chacko:2018vss,Koren:2019iuv,Feng:2020urb,Beauchesne:2020mih,Chacko:2021vin,Bansal:2021dfh,Bodas:2021fsy,Bansal:2022qbi,Ireland:2022quc,Holst:2023hff,Bodas:2024idn,Dvali:2007hz,Dvali:2009ne,Chacko:2005pe,Chacko:2005un,Chacko:2005vw,Chacko:2016kgg,Barbieri:2005ri,Craig:2013fga,Craig:2015pha,Craig:2015xla,Craig:2016lyx,GarciaGarcia:2015fol,Farina:2015uea,Prilepina:2016rlq,Barbieri:2016zxn,Berger:2016vxi,Csaki:2017spo,Elor:2018xku,Hochberg:2018vdo,Francis:2018xjd,Harigaya:2019shz,Ibe:2019ena,Dunsky:2019upk,Csaki:2019qgb,Terning:2019hgj,Johns:2020rtp,Roux:2020wkp,Ritter:2021hgu,Curtin:2021alk,Curtin:2021spx,Huang:2019obt}.
Introducing mirror sectors as atomic dark matter (ADM), realizable through broken $\mathbb{Z}_2$ symmetries, offers remarkable potential for resolving tensions in both particle physics and cosmology \cite{Kaplan:2009de,Cyr-Racine:2012tfp,Foot:2014uba,Detmold:2014qqa,Krnjaic:2014xza,Cirelli:2016rnw,Agrawal:2016quu,GarciaGarcia:2015pnn,Berlin:2022hmt,Ryan:2021dis,Blinnikov:1983gh,Ackerman:2008kmp,Feng:2009mn,Agrawal:2017pnb,Foot:2002iy,Foot:2003jt,Foot:2004pa,Foot:2004wz,Foot:2007iy,Foot:2011ve,Foot:2013vna,Foot:2014mia,Foot:2016wvj,Ciarcelluti:2004ik,Ciarcelluti:2004ip,Ciarcelluti:2008vs,Ciarcelluti:2010zz,Ciarcelluti:2012zz,Ciarcelluti:2014scd,Cudell:2014wca,Goldberg:1986nk,Fargion:2005ep,Khlopov:2005ew,Khlopov:2008ty,Khlopov:2010pq,Khlopov:2011tn,Kaplan:2011yj,Behbahani:2010xa,Cline:2012is,Cline:2013pca,Fan:2013tia,Fan:2013yva,McCullough:2013jma,Randall:2014kta,Khlopov:2014bia,Petraki:2014uza,Petraki:2016cnz,Choquette:2015mca,Hou:2011ec,Follin:2015hya,Baumann:2015rya,Buen-Abad:2022kgf,Buen-Abad:2023uva}.
Here, a mirror sector can succinctly satisfy both outlined phenomenological criteria. 
Firstly, they provide a natural link between the two sectors, elucidating the similarity in energy densities through connections in baryogenesis \cite{Farina:2016ndq,Blinov:2021mdk,Bittar:2023kdl,Alonso-Alvarez:2023bat,Hall:2021zsk,Kilic:2021zqu}.
Secondly, they inherently justify how the mirror sector contributes the required radiation and pressure supported matter preserving $\Lambda$CDM's predictive accuracy for the specified ratios \cite{Cyr-Racine:2021oal,Greene:2023cro}.

We utilize a symmetry of cosmological observables to guide our model building process, the Free-Fall, Amplitude, and Thomson (FFAT) scaling \cite{Cyr-Racine:2021oal,Ge:2022qws}.
The FFAT symmetry prescribes a way to increase the Hubble expansion rate of the Universe by proportionally adjusting the length and rate scales of the Universe.
Specifically, the following rescaling of the energy density $\rho_i$ of species $i$, the Thomson scattering rate $\dot{\kappa}$, and the amplitude of scalar fluctuations $A_{\rm s}$ by a real parameter $\lambda > 1$ leaves dimensionless cosmological observables invariant
\begin{equation}
    \{ \rho_i  \xrightarrow{\lambda} \lambda^2 \rho_i, \dot{\kappa} \xrightarrow{\lambda} \lambda \dot{\kappa}, A_{\rm s} \xrightarrow{\lambda} A_{\rm s} /\lambda^{n_{\rm{s}}-1}  \},
\end{equation}
while leading to a larger Hubble expansion rate $H\xrightarrow{\lambda} \lambda H$. 
In the above $n_{\rm s}$ is the scalar spectral index. 
Integrating the FFAT scaling into a cosmological scenario requires addressing two, nontrivial model building questions: 
\begin{itemize}
    \item What is the form of the additional energy density needed to satisfy $\rho_i \rightarrow \lambda^2\rho_i$?
    \item How do we modify the visible Thomson scattering rate to satisfy $\dot{\kappa} \rightarrow \lambda \dot{\kappa}$?
\end{itemize} 
Notably, ADM naturally satisfies the first model building criteria of the FFAT symmetry by allowing the inclusion of proportional additions of radiation and pressure supported matter while preserving the critical ratios predicted by $\Lambda$CDM.
However, as a generic feature of models seeking to alleviate the Hubble tension, the increase in $H_0$ will begin to cause issues with CMB observables as the ratio of the Thomson scattering rate to background expansion rate (the characteristic timescale of recombination) moves away from the $\Lambda$CDM predicted value.

To protect cosmic observables from alterations to the characteristic time scale of recombination, Thomson scattering rate must be proportionally scaled compared to the expansion rate as part of its second model building criteria. 
Scaling the Thomson scattering rate, which depends on electron number density and the Thomson cross section, introduces theoretical challenges. 
In this study, we utilize the strategy first suggested by Ref.~\cite{Greene:2023cro}, modifying the parameter values for the electromagnetic fine-structure constant and electron mass at the percent level.
This adjustment achieves the necessary scaling of the Thomson scattering rate while maintaining the binding energy of light elements at the Standard Model value.
As we will see, this effectively renders the CMB insensitive to these changes at first order.
However, varying fundamental constants is notoriously difficult to realize in a physical Lagrangian, for example see Refs.~\cite{Vacher:2022sro,Vacher:2024qiq}.
We point out that this method is not the only way to scale the Thomson scattering rate. 
It represents a straightforward phenomenological model that captures interactions beyond the Standard Model potentially affecting the Thomson scattering rate.
Potential interactions between the dark and visible sectors, currently beyond experimental detection limits, could be incorporated within the mirror sector framework. 
Such integration may provide a compelling `two-for-one' realization of the necessary phenomenology, effectively leveraging the unique capabilities of the new, subdominant ADM component \cite{Brzeminski:2020uhm,Dzuba:2024src}.
The relationship between fundamental constant variation and cosmology is a well discussed topic, for more models and recent discussion please see Refs.~\cite{Hart:2019dxi,Sekiguchi:2020teg,Hart:2021kad,Lee:2022gzh,Hart:2022agu,Tohfa:2023zip,Seto:2023yal,Zhang:2022ujw,Chakrabarti:2023zud,Yeung:2022smn, Colaco:2020ndf, Lopez-Honorez:2020lno, Hees:2020gda, Colaco:2023iel, Berke:2022rjk,Baryakhtar:2024rky,Seto:2024cgo}.

In Section \ref{sec:Model}, we establish the theoretical framework for the model's two key components: the $\mathbb{Z}_2$ symmetry inspired model of ADM and fundamental constant variation (FCV). 
Section \ref{sec:methodology} details our numerical investigation of this model using Monte-Carlo-Markov-Chain methods. 
We present the results in Section \ref{sec:results}, revealing a cosmological inference of $H_0$ that aligns with locally measured values.
Concluding remarks follow in Section \ref{sec:conc}. 
Throughout this paper, we adopt natural units where $c = \hbar = 1$, representing the speed of light and reduced Planck constant, respectively.
We also use the reduced Planck mass, $M_{\rm pl} = (8\pi G)^{-1/2}$, where $G$ is Newton's gravitational constant.

\section{\label{sec:Model}The Synchronous Mirror Recombination Model}

To understand how a mirror sector containing a subdominant ADM component may allow for concordant cosmology, we explain here how its parameters can be chosen to exploit the FFAT symmetry direction, ensuring alignment with CMB and LSS data.
This requires satisfying the phenomenological criteria outlined above: $\rho_i \rightarrow \lambda^2\rho_i$ and $\dot{\kappa} \rightarrow \lambda \dot{\kappa}$.
Subsection \ref{subsec:ADM} details how we fulfill the energy density requirement to increase $H$, and subsection \ref{subsec:FCV} describes the scaling of the Thomson scattering rate to ensure consistent recombination history.

\subsection{\label{subsec:ADM}Atomic Dark Matter}

Mirror-like dark sectors can vary widely, ranging from those featuring multiple, exact replicas of the Standard Model to others with only partial duplications \cite{Arkani-Hamed:2016rle,Chacko:2016hvu,Chacko:2018vss,Koren:2019iuv,Feng:2020urb,Beauchesne:2020mih,Chacko:2021vin,Bansal:2021dfh,Bodas:2021fsy,Bansal:2022qbi,Ireland:2022quc,Holst:2023hff,Bodas:2024idn,Dvali:2007hz,Dvali:2009ne,Chacko:2005pe,Chacko:2005un,Chacko:2005vw,Chacko:2016kgg,Barbieri:2005ri,Craig:2013fga,Craig:2015pha,Craig:2015xla,Craig:2016lyx,GarciaGarcia:2015fol,Farina:2015uea,Prilepina:2016rlq,Barbieri:2016zxn,Berger:2016vxi,Csaki:2017spo,Elor:2018xku,Hochberg:2018vdo,Francis:2018xjd,Harigaya:2019shz,Ibe:2019ena,Dunsky:2019upk,Csaki:2019qgb,Terning:2019hgj,Johns:2020rtp,Roux:2020wkp,Ritter:2021hgu,Curtin:2021alk,Curtin:2021spx}. 
Importantly, these frameworks often allow for masses and coupling strengths of mirror particles to be comparable in magnitude to the visible sector. 
Given the variety and scope of models exploring ADM and $\mathbb{Z}_2$ symmetries, our approach in this study is broad, aiming to delineate dark sector phenomenology that could be compatible with various mirror ADM implementations \cite{Kaplan:2009de,Cyr-Racine:2012tfp,Foot:2014uba,Detmold:2014qqa,Krnjaic:2014xza,Cirelli:2016rnw,Agrawal:2016quu,GarciaGarcia:2015pnn,Berlin:2022hmt,Ryan:2021dis,Blinnikov:1983gh,Ackerman:2008kmp,Feng:2009mn,Agrawal:2017pnb,Foot:2002iy,Foot:2003jt,Foot:2004pa,Foot:2004wz,Foot:2007iy,Foot:2011ve,Foot:2013vna,Foot:2014mia,Foot:2016wvj,Ciarcelluti:2004ik,Ciarcelluti:2004ip,Ciarcelluti:2008vs,Ciarcelluti:2010zz,Ciarcelluti:2012zz,Ciarcelluti:2014scd,Cudell:2014wca,Goldberg:1986nk,Fargion:2005ep,Khlopov:2005ew,Khlopov:2008ty,Khlopov:2010pq,Khlopov:2011tn,Kaplan:2011yj,Behbahani:2010xa,Cline:2012is,Cline:2013pca,Fan:2013tia,Fan:2013yva,McCullough:2013jma,Randall:2014kta,Khlopov:2014bia,Petraki:2014uza,Petraki:2016cnz,Choquette:2015mca,Hou:2011ec,Follin:2015hya,Baumann:2015rya}.

We assume the mirror sector primarily consists of two massive stable particles, each bearing a charge under a dark $U(1)$ symmetry reminiscent of the visible electromagnetic sector.
For the sake of simplicity in modeling, these particles are assumed to be fundamental fermions, though we recognize that composite particles could serve equally well in a more specific realization. 
These fermions are distinguished by their masses: one mirroring the mass of the visible electron and the other mirroring the mass of the visible proton. 
We refer to the lighter of these as the dark electron and the heavier as the dark proton, with their interactions mediated by a massless dark photon introducing a dark radiation component to the Universe. 
These additions to the dark sector interact with the visible sector exclusively through gravitational forces and represent a subdominant portion of the overall dark sector energy density.
For the scope of this work, we do not define the specific characteristics of the remaining dark matter, aside from its adherence to CDM behavior. 
However, we note an interesting possibility and connection to recent research: the CDM could for example be a stable `dark neutron' left over from a dark nucleosynthesis era in the early Universe which may naturally agree with our ADM approach \cite{Bodas:2024idn}.

This schematic model presents three significant phenomenological implications. 
Firstly, it enables the formation of bound states akin to visible baryons.
Secondly, the capacity to form bound states means the mirror sector will undergo a recombination process analogous to that experienced by the visible sector. 
Thirdly, ADM models allow the critical ratios established by $\Lambda$CDM to be preserved:
\begin{itemize}
    \item The precisely measured baryon-photon ratio is preserved through the proportional inclusion of ADM and dark radiation.
    \item The matter-radiation ratio is maintained through the proportional inclusion of CDM, ADM, neutrinos, and dark radiation.
    \item Dark radiation lends pressure support to ADM, and the integration of extra cold dark matter ensures a balance between pressure-supported and pressureless matter.
    \item At appropriate temperatures, radiation interacting with ADM behaves as a fluid, maintaining this state up to `mirror recombination', akin to the recombination process in the visible sector. Following this, the dark radiation shifts to freestreaming, thereby maintaining the fluid-freestreaming radiation ratio when the recombination events align.
\end{itemize}
Models of atomic dark matter, especially those derived from partial $\mathbb{Z}_2$ symmetries, naturally define a parameter space encompassing mirror recombination which occur near visible recombination. 
The proportional inclusion of the mirror sector densities is the same statement as the second clue outlined above.
It is natural that the ratios of dark sector energies matches that of the visible sector if they are populated by the same criteria.
Together, this approach circumvents numerous fine-tuning challenges.

\subsection{Thermal History of Mirror Sector}
To limit the number of free parameters of the theory during our numerical investigation, we fix the dark proton mass $m_{\rm p,d}$, dark electron mass $m_{\rm e,d}$, and dark fine structure constant $\alpha_{\rm d}$ to be functions of the mirror sector temperature $\xi_{\rm d}$ such that mirror recombination occurs at the same time as the visible sector, ensuring the fluid-freestreaming radiation ratio is preserved.
The fraction of total dark matter which is atomic, $f_{\rm ADM}$, is left as a free parameter.
We call this scenario the `synchronous mirror recombination' case.
Later, in section \ref{sec:brokenmirror} we will consider a model where the mirror sector is allowed to recombine independently of the visible sector in the `asynchornous mirror recombination' case.
To ensure synchronous mirror recombination and enforce the fourth condition regarding the fluid-freestreaming radiation ratio, we make three assumptions regarding the mirror sector: (1) $\dot{\kappa}$ is similar in both sectors, (2) the sound speeds in both the sectors are similar to first order, and (3) the ratio of binding energy to temperature is congruent between the two sectors. 
The first constraint ensures dark photons appropriately couple to dark electrons. 
The second constraint guarantees that pressure waves in the mirror plasma propagate at speeds matching those in the visible sector; otherwise, differing speeds of dark pressure waves would gravitationally affect visible sector pressure waves, altering their profile slightly. 
The third constraint determines the conditions under which recombination in the mirror sector becomes energetically favorable.

The first assumption is that the Thomson scattering rates for both sectors is the same:
\begin{equation} \label{eq:scatter}
    \lambda n_{\rm e} a \sigma_{\rm T} = n_{\rm e,d} a \sigma_{\rm T,d}.
\end{equation}
where $\lambda = \Tilde{H}_0/H_{\rm 0, \Lambda CDM}$ with $\Tilde{H}_0$ being the increased value and the subscript $d$ denotes dark sector values.
This formulation introduces a new variable; the dark electron number density. 
Similar to the visible sector, we account for this by assuming global neutrality under the dark $U(1)$ symmetry, leading to the relation
\begin{equation} \label{eq:darknuetral}
    n_{\rm e,d} = n_{\rm p,d} \approx \frac{\rho_{\rm ADM}}{m_{\rm p,d}}
\end{equation}
where $n_{\rm p,d}$ and $\rho_{\rm ADM}$ represent the dark proton number density and ADM energy density, respectively. 
Therefore, the mass of the dark proton emerges as a key determinant of the dark electron number density. 
Equation \eqref{eq:darknuetral} holds true specifically when $m_{\rm e,d} \ll m_{\rm p,d}$, a condition inherently met in mirror-like models.

From the degeneracy direction pointed out by the FFAT symmetry, the visible baryon density is related to the ADM density such that
\begin{equation}    
    \rho_{\rm ADM} = \left(\lambda^2-1\right)\rho_{\rm b}.
\end{equation}
Therefore, equation \eqref{eq:scatter} becomes
\begin{equation}
    \lambda \frac{\rho_{\rm b}}{m_{\rm p}} \sigma_{\rm T} = \frac{\left(\lambda^2-1\right)\rho_{\rm b}}{m_{\rm p,d}} \sigma_{\rm T,d}.
\end{equation}
For clarity, we define the ratio of a quantity $\chi$ and its dark sector component $\chi_{\rm d}$ as $r_{\chi} = \frac{\chi_{\rm d}}{\chi}$.
With this notation, the constraint from \eqref{eq:scatter} becomes
\begin{equation} \label{eq:scatter4}
    \frac{r_{\rm mp}r_{\rm me}^2}{r_{\alpha}^2} = \frac{\xi_{\rm d}^4}{\sqrt{\xi_{\rm d}^4+1}}
\end{equation}
where $\alpha$ is the visible fine-structure constant and we have made the substitution
\begin{equation}
    \xi_{\rm d} \equiv \frac{T_{\rm \gamma,d}}{T_{\rm \gamma}} =  \left(\lambda^2 - 1\right)^{1/4}
\end{equation}
with $T_{\rm \gamma}$ being the visible sector temperature and $T_{\rm \gamma,d}$ being the mirror sector temperature.

We now seek to write $r_{\rm p}$ in terms of $\xi_{\rm d}$ and can do so while enforcing the second assumption.
The sound speed of the primordial plasma is given by
\begin{equation}
    c_{\rm s}^2 = \frac{T_{\rm b}}{\bar{\mu}}\left(1 - \frac{1}{3}\frac{d{\rm ln}T_{\rm b}}{d{\rm ln}a}\right)
\end{equation}
where $T_{\rm b}$ is the temperature of baryons and $\bar{\mu}$ is the reduced molecular weight.
At first order, we can ensure $c_{\rm s} = c_{\rm s,d}$ by enforcing
\begin{equation} \label{eq:soundspeed}
    \frac{T_{\rm b}}{\bar{\mu}} = \frac{T_{\rm ADM}}{\bar{\mu}_{\rm d}}
\end{equation}
where $T_{\rm ADM}$ is the temperature of the atomic dark matter.
Prior to recombination, the baryons and ADM are tightly coupled to their respective radiation baths such that $T_{\rm b} \approx T_{\gamma}$ and $T_{\rm ADM} \approx T_{\gamma, d}$.
Additionally, $\bar{\mu} \approx m_{\rm p}$ as long as $m_{\rm p} \gg m_{\rm e}$.
Therefore, \eqref{eq:soundspeed} implies
\begin{equation}
\begin{split}
    \frac{T_{\rm ADM}}{T_{\rm b}} \approx \frac{T_{\rm \gamma,d}}{T_{\gamma}} &=  \frac{m_{\rm p,d}}{m_{\rm p}} \\
    \Rightarrow \xi_{\rm d} &= r_{\rm mp}.
\end{split}
\end{equation}
The constraint on the dark electron mass and dark fine structure constant from equation \eqref{eq:scatter4} becomes 
\begin{equation}
        \frac{r_{\rm me}^2}{r_{\alpha}^2} = \frac{\xi_{\rm d}^3}{\sqrt{\xi_{\rm d}^4+1}} .
\end{equation}
We now turn our attention to the last and perhaps most important constraint: keeping the ratio of the binding energy $B_{\rm e}$ to temperature in both sectors equal such that
\begin{equation}
    \frac{B_{\rm e}}{T_{\gamma}} = \frac{B_{\rm e,d}}{T_{\rm \gamma,d}}
\end{equation}
which simplifies to the constraint
\begin{equation}
    \xi_{\rm d} = r_{\rm e}r_{\alpha}^2
\end{equation}
where we have used $B_{\rm e(,d)} = \frac{1}{2}m_{\rm e(,d)}\alpha_{\rm (d)}^2$.
Therefore, the system of equations we must solve for $r_{\rm e}$ and $r_{\alpha}$ is
\begin{equation}
    r_{\rm me}r_{\alpha}^2 = \xi_{\rm d} \And \frac{r_{\rm me}^2}{r_{\alpha}^2} = \frac{\xi_{\rm d}^3}{\sqrt{\xi_{\rm d}^4+1}}.
\end{equation}
We find the mirror sector parameters to enforce synchronous mirror recombination to be solely functions of temperature such that
\begin{equation}\label{eq:fixed_ds_params}
\begin{split}
    r_{\rm me} &= \left(\frac{\xi_{\rm d}^4}{\sqrt{\xi_{\rm d}^4+1}}\right)^{1/3} \\
    r_{\rm mp} &= \xi_{\rm d} \\
    r_{\alpha} &= \left(\frac{\sqrt{\xi_{\rm d}^4+1}}{\xi_{\rm d}}\right)^{1/6}.
\end{split}
\end{equation}
This approach ensures that the forth condition of preserving the fluid-freestreaming radiation ratio is exactly preserved.
However, we note that the mirror sector's subdominant nature offers flexibility in these relationships; exact matches are not required. 
We explore this flexibility when we permit the mirror sector to recombine independently from the visible sector in section \ref{sec:brokenmirror}.
We note that this parameterization begins to break down for very cold mirror sectors where $\xi_{\rm d} \rightarrow 0$, as not only do the masses and couplings diverge, but the introduced ADM will be cold and behave as CDM which we describe in section \ref{sec:coldthumbs}.

\subsection{\label{subsec:FCV}Phenomenological Changes to $\dot{\kappa}$ using FCV}
The dimensionless ratio of the Thomson scattering rate, $\dot{\kappa}$, to the Hubble rate, $H(z)$, is crucial for setting the photon diffusion scale during recombination. 
This period, vital for shaping the CMB spectra at small angular scales, demands a balance between the expansion and Thomson scattering rates for observational consistency. 
The high $\ell$ regions of the CMB spectra, where $\dot{\kappa}$ smooths perturbations resulting in power dampening, are particularly sensitive to this ratio. 
Thus, the Hubble tension can be reframed as a `photon diffusion tension' in this context \cite{Greene:2023cro}.

The Thomson cross section dictates the mean free path of photons at this time, and variations in $\alpha$ and $m_{\rm e}$ can enhance or hinder the level of diffusion. 
Our method employs a targeted FCV approach, focusing on maintaining constant binding energy and $\dot{\kappa}/H(z)$, the physical quantities which drives observational constraints. 
In essence, CMB observations are more attuned to changes in the binding energy and this ratio, rather than specific alterations in $\alpha$ or $m_{\rm e}$.
Notably, CMB \textit{spectral} observations may be sensitive to specific variations in $\alpha$ or $m_{\rm e}$ with possible future telescopes but remain out of reach for current observations \cite{Fixsen:1996nj,Chluba:2019nxa, Chluba:2022xsd,Chluba:2022efq,Kite:2022eye}.
We remind the reader that this phenomenological approach encapsulates any modification to the Thomson scattering rate potentially achievable through novel interactions between the dark and visible sectors, and is not solely limited to models investigating FCV.

Our proposed scaling for $\dot{\kappa}$ hinges on two constraints. 
Firstly, the binding energy of light elements $B_{\rm e} =\frac{1}{2} m_{\rm e}\alpha^2$ must remain unchanged compared to the Standard Model to ensure visible recombination timing aligns with $\Lambda$CDM.
This preserves observables related to the distance to the CMB.
Secondly, the Thomson cross section $\sigma_{\rm T} = \frac{8\pi}{3}\alpha^2/m_{\rm e}^2$ requires a specific scaling ($\sigma_{\rm T} \rightarrow \lambda \sigma_{\rm T}$) to maintain the $\dot{\kappa}/H$ ratio, thereby preserving CMB temperature spectra observational consistency. 
Resolving these constraints yields precise scaling predictions for the fundamental parameters: $\alpha \rightarrow \lambda^{1/6} \alpha \And m_{\rm{e}} \rightarrow \lambda^{-1/3} m_{\rm{e}}$ \cite{Ge:2022qws,Greene:2023cro}. 
These scalings align with the degeneracy direction indicated by the FFAT symmetry, while preserving recombination-related cosmological observables at first order.
For more discussion regarding the observational constraints imposed on FCV, see App.~\ref{ap:FCVconstraints}, and for results depicting the relationship between FCV and the binding energy see Sec.~\ref{subsec:FCV}.

\section{\label{sec:methodology} Methodology}
We integrate the ADM model into the Cosmology Linear Anisotropy Solving System (\texttt{CLASS}) to probe the mirror sector parameter space, and allow the prior volume to include the FFAT symmetry direction \cite{Blas:2011rf}.
To incorporate the FCV necessary to keep $\dot{\kappa}/H$ invariant, we utilize \texttt{CLASS}'s default functionality for varying $\alpha$ and $m_{\rm e}$ and call upon \texttt{HyREC} to compute recombination for the mirror and visible sectors \cite{Lee:2020obi}.
Combining \texttt{CLASS} with the Monte-Carlo-Markov-Chain (MCMC) package \texttt{MontePython} \cite{Audren:2012wb,Brinckmann:2018cvx}, we compute the one and two-dimensional posteriors for the free parameters for the synchronous mirror recombination model.
To plot our derived 1 and 2 dimensional posteriors, we utilize the python package \texttt{GetDist} \cite{Lewis:2019xzd}.

Although this approach introduces a considerable expansion in the model's degrees of freedom, they are well justified by physical principles and guided by underlying symmetries.
For clarity, we list the free parameters incorporated into the model, along with their physical justification.
First, we consider the six standard $\Lambda$CDM parameters $\left[h,\omega_{\rm b},\omega_{\rm cdm},A_{\rm s},n_{\rm s},\tau_{\rm reio}\right]$.
To maintain a constant ratio of fluid-freestreaming radiation, we introduce a variable neutrino energy density $\left[N_{\rm ur}\right]$ while assuming that neutrinos are massless. 
Additionally, we incorporate variations in the fundamental parameters $\left[\alpha, m_{\rm e}\right]$ such that the Thomson scattering rate can vary while keeping the binding energy constant.
We allow $\left[f_{\rm ADM},\xi_{\rm d}\right]$ to vary, enabling the mirror sector to preserve the matter-radiation, baryon-photon, and pressure supported to pressureless matter ratios.
The other mirror sector parameters ($m_{\rm p,d}, m_{\rm p,e}, \alpha_{\rm d}$) are fixed functions of the temperature according to Eq.~\eqref{eq:fixed_ds_params}, ensuring a synchronous mirror recombination with the visible sector.

We consider the following observational data sets
\begin{itemize}
    \item CMB - Planck 2018 high $\ell$ TT, TE, and EE data, Planck 2018 low $\ell$ TT and EE data, and Planck Lensing data \cite{Planck:2018vyg}
    \item Baryon Acoustic Oscillations (BAO) - 6dF Galaxy Survey \cite{Beutler:2011hx}, SDSS DR7 \cite{Ross:2014qpa}, and BOSS DR12 \cite{BOSS:2016wmc}
    \item Uncalibrated Pantheon+ Supernova \cite{Brout:2022vxf}
    \item Pantheon+ Supernova calibrated by the local S$H_0$ES measurement \cite{Riess:2021jrx}
\end{itemize}
in three different combinations: (A) CMB and BAO, (B) CMB, BAO, and uncalibrated Pantheon+, and (C) CMB, BAO, and Pantheon+ calibrated by S$H_0$ES, referred to hereafter as likelihood groups A, B, and C.
In all cases, we set the helium abundance to $Y_{\rm p} = 0.2454$ to ensure alignment with BBN predictions \cite{Aver:2020fon}.
As mentioned previously, models where the dark sector thermalizes with the visible sector after BBN, thereby avoiding $N_{\rm eff}$ constraints \cite{Berlin:2019pbq,Berlin:2022hmt}, present intriguing connections and are certainly possible, but are beyond the scope of this work.

The MCMC analysis is allowed to run until the chains satisfy a Gelman-Rubin convergence criterion of $R-1 < 0.05$ in all free parameters.
To identify the global minimum of the fits (a significant computational challenge given the extensive parameter space spanned by likelihood groups A and B as we will see) we employ the newly developed \texttt{Procoli} package which comes equipped with an integrated global best-fit minimizer \cite{Karwal:2024qpt}.
\texttt{Procoli} is also utilized for conducting profile likelihood scans, enabling us to understand the relationship between the fit to data and increasing values of $H_0$ when exploring the FFAT symmetry.
Additionally, we define the visible sector variables
\begin{equation}
    \begin{split}
        \delta_{\alpha} =&
        \frac{\Tilde{\alpha}-\alpha_0}{\alpha_0} \\
        \delta_{\rm e} =& 
        \frac{\Tilde{m}_{\rm e}-m_{\rm e,0}}{m_{\rm e,0}} 
    \end{split}
\end{equation}
where the tilde indicates the varied value and 0 indicates the Standard Model value.

\section{\label{sec:results} Results and Discussion}
\begin{figure*}
\includegraphics[width=\textwidth]{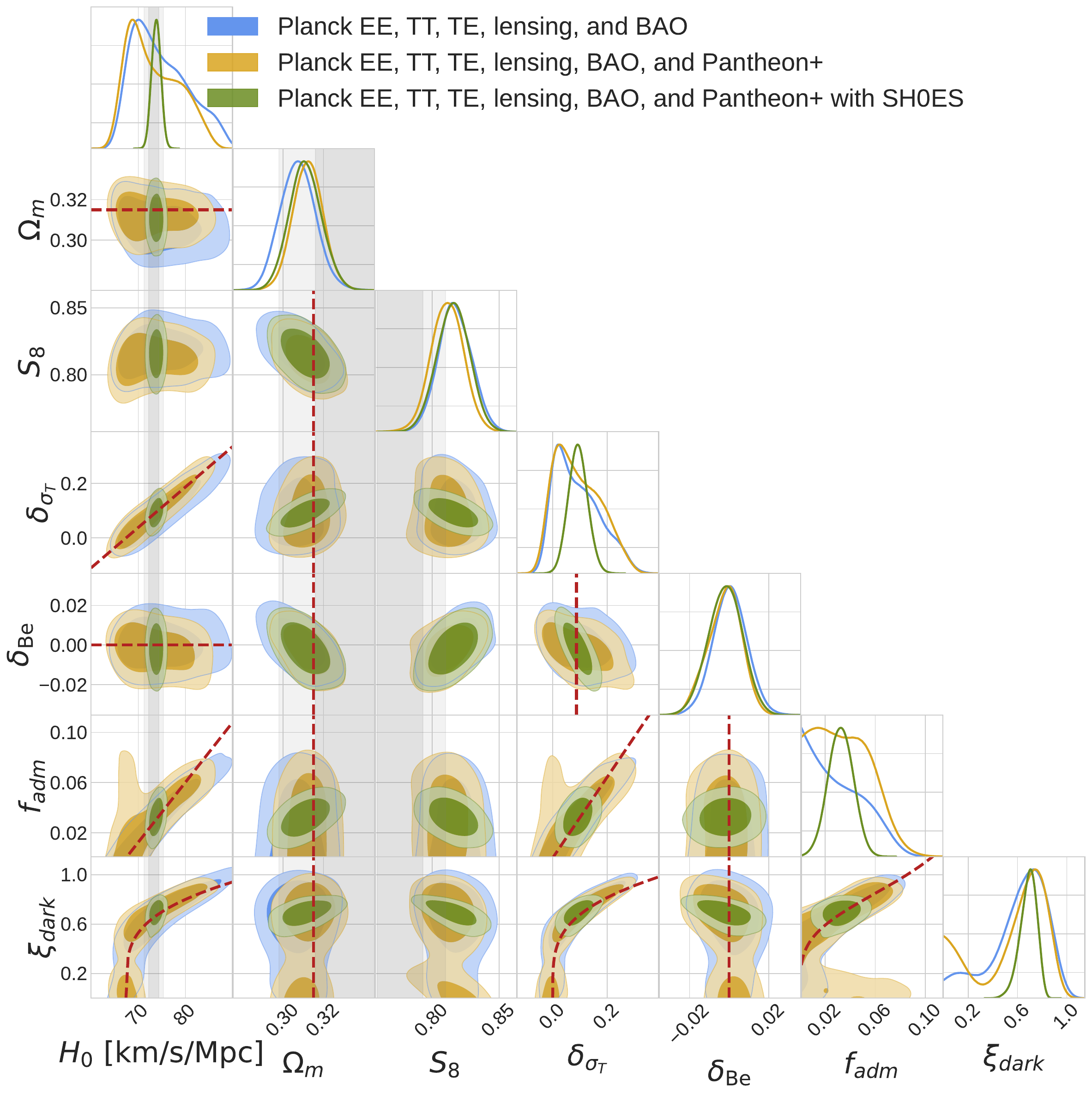}
\centering
\caption{This triangle plot showcases the MCMC analysis results for the synchronous recombination mirror model under three different likelihood groups: A (blue), B (yellow), and C (green). Dark red dashed lines represent the degeneracy direction of the FFAT symmetry. Note that the FFAT symmetry does not make a prediction for $S_8$. In the $H_0$ column, dark and light grey bands mark the 68\% and 95\% CL from the S$H_0$ES' locally measured value of $H_0 = 73.3 \pm 1.04$ \cite{Riess:2021jrx}, while similar bands in the $\Omega_{\rm m}$ and $S_8$ columns reflect the constraints from Pantheon+'s measured value of $\Omega_{\rm m} = 0.334 \pm 0.018$ \cite{Brout:2022vxf} and DES-Y3 result of $S_8 = 0.776\pm0.017$ \cite{DES:2021wwk}. The MCMC sampler naturally follows the FFAT symmetry direction as it ensures agreement with CMB and LSS observational data. The degeneracy is broken only once the local S$H_0$ES calibration of $H_0$ is included, effectively selecting a slice of the posterior.}
\label{fig:threecontour}
\end{figure*}

We tabulate the results for the scans over the synchronous mirror recombination model in table \ref{tab:my_label} and show the posteriors in figure \ref{fig:threecontour} for likelihood groups A, B, and C.
Additionally, we perform the same analysis for the $\Lambda$CDM model using the same likelihoods to generate comparative data.
In general, we find that all parameters naturally follow the FFAT symmetry direction, depicted by red dashed lines, and keep the critical ratios discussed above invariant compared to $\Lambda$CDM.
The binding energy is left invariant compared to the Standard Model while the Thomson cross section is scaled to match the change in $H$ as expected.

\begin{table*}
    \centering
    \begin{tabular}{ c  c  c  c  c }
    \hline
        & Free Parameters & CMB $\And$ BAO & CMB, BAO, Pantheon+ & CMB, BAO, Pantheon+, S$H_0$ES \\
    \hline
    $\Lambda$CDM    & $h$   & $\hspace{5 mm}0.7462\hspace{1 mm}(0.6854)\pm0.0549\hspace{5 mm}$    
                            & $\hspace{5 mm}0.7273\hspace{1 mm}(0.7370)\pm0.0549\hspace{5 mm}$ 
                            & $\hspace{5 mm}0.7380\hspace{1 mm}(0.7380)\pm0.0102\hspace{5 mm}$ \\
        & $\omega_{\rm b}$    & $0.0223\hspace{1 mm}(0.0224)\pm0.0002$
                            & $0.0223\hspace{1 mm}(0.0224)\pm0.0002$
                            & $0.0223\hspace{1 mm}(0.0226)\pm0.0002$\\
        & $\omega_{\rm cdm}$  & $0.1495\hspace{1 mm}(0.1221)\pm0.0244$        
                            & $0.1439\hspace{1 mm}(0.1475)\pm0.0256$ 
                            & $0.1472\hspace{1 mm}(0.1473)\pm0.0065$ \\
        & $n_{\rm s}$         & $0.9585\hspace{1 mm}(0.9580)\pm0.0095$
                            & $0.9574\hspace{1 mm}(0.9568)\pm0.0089$
                            & $0.9585\hspace{1 mm}(0.9589)\pm0.0096$\\
        & $A_{\rm s}10^9$ & $2.0794\hspace{1 mm}(2.0740)\pm0.0411$
                            & $2.0728\hspace{1 mm}(2.0736)\pm0.0368$
                            & $2.0792\hspace{1 mm}(2.0802)\pm0.0417$\\
        & $\tau_{\rm reio}$   & $0.0518\hspace{1 mm}(0.0550)\pm0.0077$
                            & $0.0520\hspace{1 mm}(0.0512)\pm0.0072$
                            & $0.0521\hspace{1 mm}(0.0516)\pm0.0075$\\
    \hline
    ADM    & $\xi_{\rm dark}$    & $0.6242\hspace{1 mm}(0.4989)\pm0.1954$          
                            & $0.5615\hspace{1 mm}(0.6977)\pm0.2660$ 
                            & $0.6849\hspace{1 mm}(0.6893)\pm0.0707$ \\
           &$f_{\rm adm}$       & $0.0317\hspace{1 mm}(0.0072)\pm0.0212$ 
                            & $0.0359\hspace{1 mm}(0.0340)\pm0.0212$ 
                            & $0.0322\hspace{1 mm}(0.0346)\pm0.0103$ \\
    \hline
    FCV    &$\delta_{\alpha}$   & $0.0142\hspace{1 mm}(0.0021)\pm0.0120$         
                            & $0.0122\hspace{1 mm}(0.0152)\pm0.0118$ 
                            & $0.0141\hspace{1 mm}(0.0149)\pm0.0040$ \\
        &$\delta_{\rm me}$  & $-0.0264\hspace{1 mm}(-0.0014)\pm0.0264$        
                            & $-0.0256\hspace{1 mm}(-0.0309)\pm0.0260$ 
                            & $-0.0293\hspace{1 mm}(-0.0296)\pm0.0132$ \\
    \hline
    Other   & $N_{\rm ur}$        & $3.3860\hspace{1 mm}(2.8063)\pm0.5624$            
                            & $3.2670\hspace{1 mm}(3.3393)\pm0.5370$ 
                            & $3.3713\hspace{1 mm}(3.3793)\pm0.3481$ \\
    \hline
    &Derived Parameters &&&\\
    \hline
    Dark &$r_{\alpha}$& $1.1291\hspace{1 mm}(1.1286)\pm0.1305$
                            & $1.1705\hspace{1 mm}(1.0808)\pm0.2130$
                            & $1.0816\hspace{1 mm}(1.0822)\pm0.0183$\\
        &$r_{\rm me}$& $0.5291\hspace{1 mm}(0.3917)\pm0.2201$
                            & $0.4996\hspace{1 mm}(0.5973)\pm0.2524$
                            & $0.5906\hspace{1 mm}(0.5886)\pm0.0774$\\
        &$r_{\rm mp}$& $0.6242\hspace{1 mm}(0.4989)\pm0.0195$
                            & $0.5615\hspace{1 mm}(0.6977)\pm0.2660$ 
                            & $0.6849\hspace{1 mm}(0.6893)\pm0.0707$ \\
     \hline
     Visible&$\delta_{\rm Be}$    & $0.0004\hspace{1 mm}(0.0028)\pm0.0086$
                            & $-0.0023\hspace{1 mm}(-0.0012)\pm0.0085$
                            & $-0.0020\hspace{1 mm}(-0.0005)\pm0.0087$\\
        &$\delta_{\sigma_{T}}$    & $0.0901\hspace{1 mm}(0.0070)\pm0.0831$
                            & $0.0903\hspace{1 mm}(0.0974)\pm0.0839$
                            & $0.0929\hspace{1 mm}(0.0938)\pm0.0355$\\
        &$\Omega_{\rm m}$   & $0.3070\hspace{1 mm}(0.3075)\pm0.0089$
                            & $0.3124\hspace{1 mm}(0.3123)\pm0.0077$
                            & $0.3111\hspace{1 mm}(0.3114)\pm0.0082$\\
        &$S_8$         & $0.8178\hspace{1 mm}(0.8167)\pm0.0127$
                            & $0.8113\hspace{1 mm}(0.8146)\pm0.0125$
                            & $0.8164\hspace{1 mm}(0.8159)\pm0.0128$\\
    \hline
    \hline
        &$\Delta \chi^2$    & $-1.08$ 
                            & $-1.26$ 
                            & $-35.72$ \\
    \hline
    \end{tabular}
    \caption{This table presents the mean (best-fit) values for the synchronous mirror recombination model across three different likelihood groups, with all errors representing the 68\% CL. A double horizontal line divides the table into two main sections. The upper section lists free parameters, categorized into $\Lambda$CDM, atomic dark matter (ADM), and fundamental constant variation (FCV). The lower section details derived parameters from both the dark and visible sectors. At the bottom, the table features $\Delta \chi^2$ values, comparing the synchronous mirror recombination model's $\chi^2$ to a baseline $\Lambda$CDM model using identical likelihoods.}
    \label{tab:my_label}
\end{table*}

Table \ref{tab:my_label} shows that without Pantheon+ supernova data and S$H_0$ES calibration, likelihood group A best-fit values favour $\Lambda$CDM model values.
This matches our clues discussed in Sec.~\ref{sec:intro}: without a need to introduce an additional signal into CMB data to increase $H_0$, $\Lambda$CDM out preforms other models as it accurately predicts the ratios dictating CMB spectra with minimal parameters.
Yet, posteriors are significantly expanded, allowing for $H_0$ values well over 80 km/s/Mpc within the 68\% confidence level - a notable departure from traditional $\Lambda$CDM CMB and LSS analysis.
Figure \ref{fig:threecontour} demonstrates an expanded $H_0$ posterior which opens the parameter space to explore cosmologically concordant models.
Indeed, as we will see in our discussion of the profile likelihood for $H_0$, CMB and LSS observations have little constraining power in the FFAT symmetry direction.
The minimizer finds best-fit values consistent with $\Lambda$CDM underscoring $\Lambda$CDM's efficacy at matching observations, while also highlighting the need for cosmological models to adhere to its foundational ratios (photon-baryon number, matter-radiation ratio, pressure supported to pressureless matter ratio, fluid-freestreaming radiation ratio) for stronger statistical agreement.
Adding five parameters to the model marginally improves the fit to CMB and LSS data alone with a $\Delta \chi^2 = -1.0799$.

\begin{figure*}
\centering
    \begin{subfigure}[b]{0.49\textwidth}
        \centering
        \includegraphics[width=\textwidth]{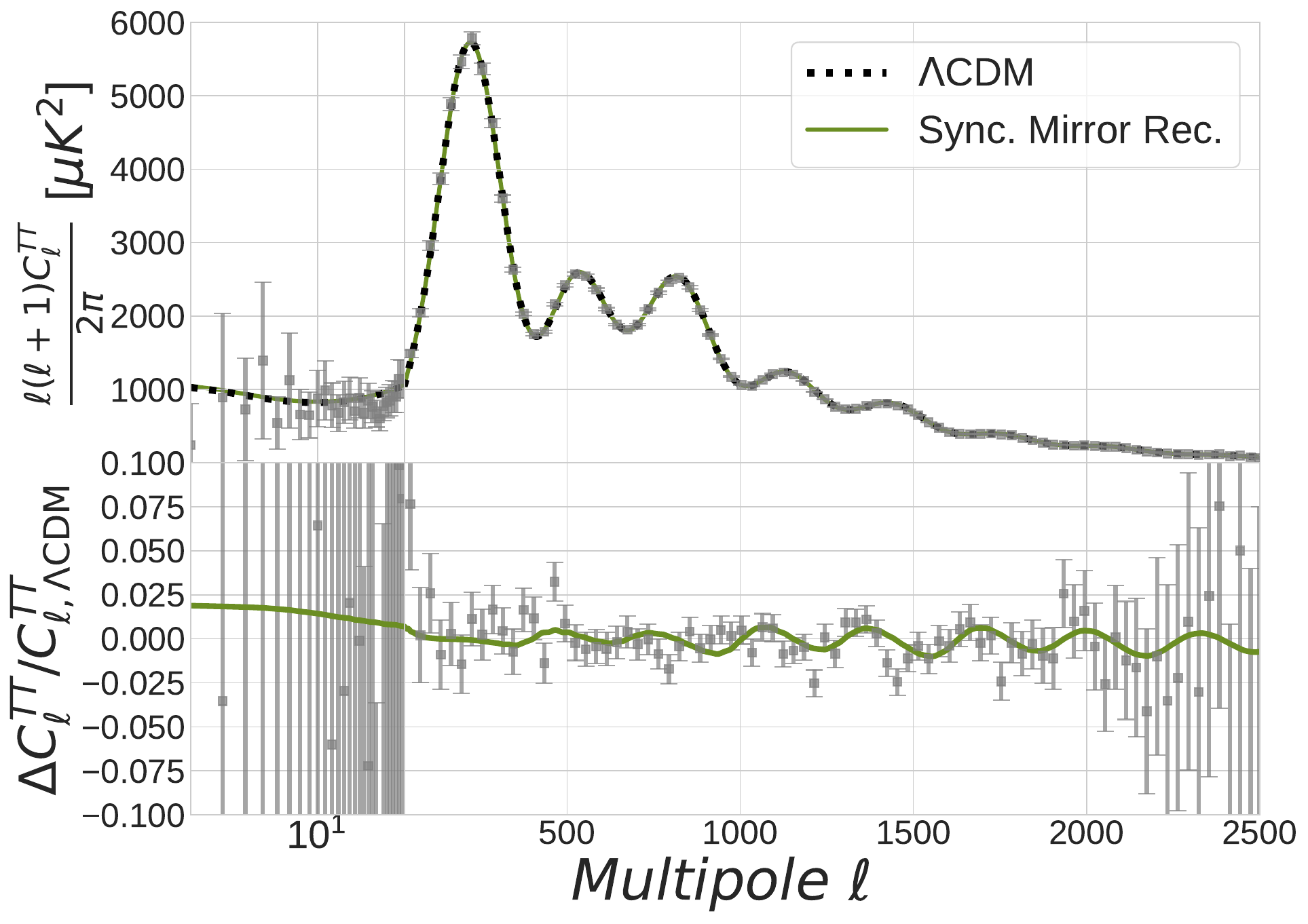}
        \caption{CMB TT Correlation Spectra}
        \label{subfig:TTspectra}
    \end{subfigure}
    \hfill
        \begin{subfigure}[b]{0.49\textwidth}
        \centering
        \includegraphics[width=\textwidth]{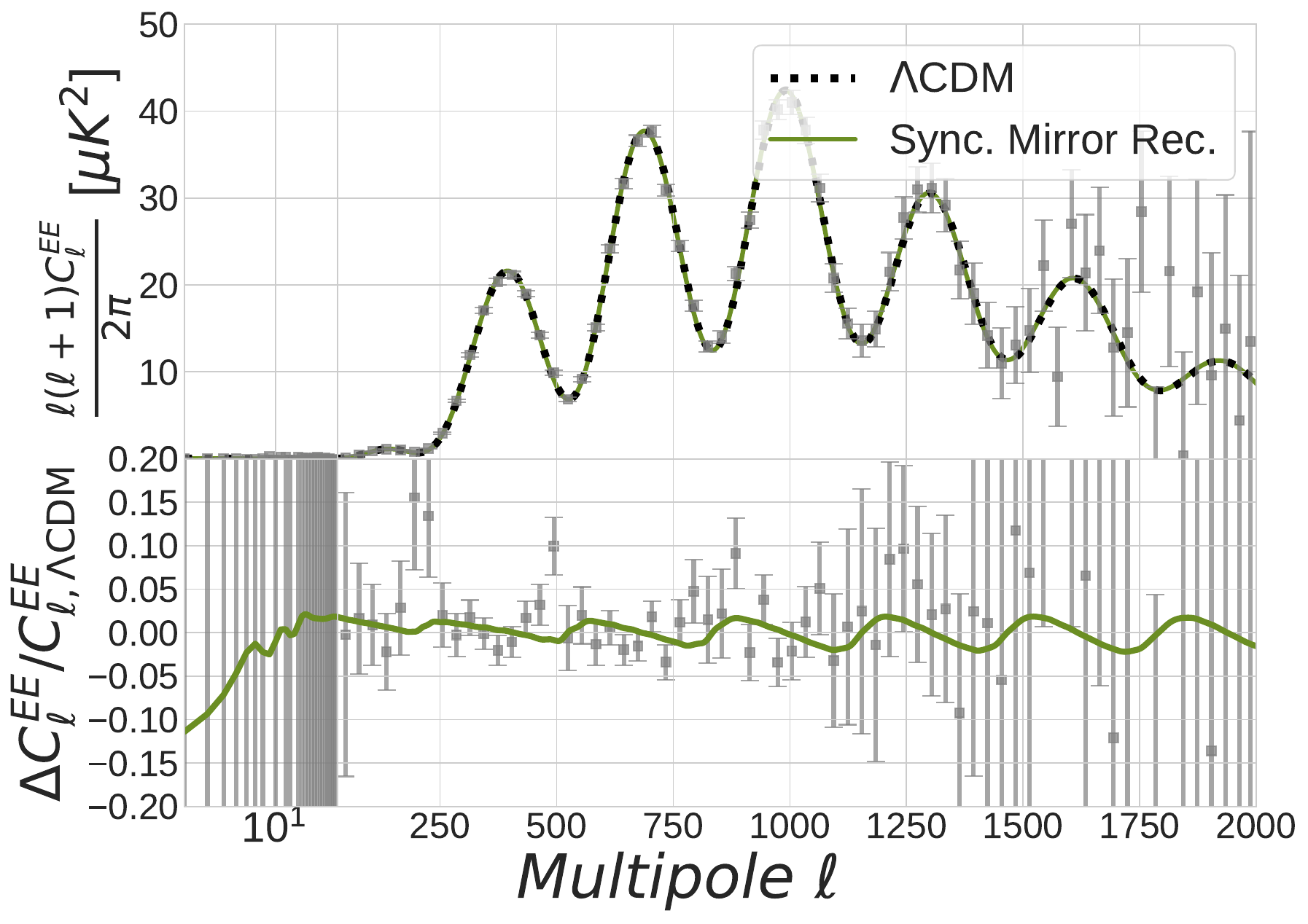}
        \caption{CMB EE Correlation Spectra}
        \label{subfig:EEspectra}
    \end{subfigure}
\centering
\caption{These two figures compare the TT and EE spectra from the $\Lambda$CDM model best-fit values obtained from Ref.~\cite{Planck:2018vyg} to the synchronous mirror recombination model best-fit values obtained when considering likelihood group C. In both figures, $H_0$ = 73.8 km/s/Mpc. The top panels of the figures demonstrate the spectra's power correlation. The bottom panels demonstrate the fractional difference between the $\Lambda$CDM and synchronous model. The grey points with associated error bars are binned Planck 2018 observations of the TT and EE spectra. The plots are logarithmic on the x axis for $2 \leq \ell < 30$ for the low $\ell$ data, and then linear for $\ell \geq 30$ for the high $\ell$ data.}
\label{fig:spectra}
\end{figure*}
When incorporating the uncalibrated Pantheon+ supernovae data into the analysis, the best-fit preferences for the model shift dramatically, naturally leading to a best-fit $H_0 = 73.70$ km/s/Mpc.
The inclusion of uncalibrated supernova data makes the model sensitive to the shape of the redshift-luminosity relationship and probes the eras of matter domination and matter-dark energy equality, but in this case is unaware of the S$H_0$ES calibration.
Notably, the best-fits agrees nicely with direct local measurements of the expansion rate without notion of their existence.
For both likelihood groups A and B, the error bars are broad because the sampler is traversing the FFAT symmetry direction without a calibration for $H_0$.
Although the best-fits now align with local measurements of $H_0$, the enhancement to the overall fit remains minimal, with a $\Delta \chi^2 = -1.2602$.
The two-dimensional posteriors for likelihood groups A and B in figure \ref{fig:threecontour} align well with our predictions from the FFAT symmetry, indicated by the red dashed lines.

The inclusion of the S$H_0$ES' local $H_0$ measurement into likelihood group C calibrates the FFAT symmetry, significantly reducing parameter uncertainties and anchoring the results.
With S$H_0$ES calibration, we find $H_0 = 73.80 \pm 1.02$ km/s/Mpc, closely matching the local measurement of Ref.~\cite{Riess:2021jrx} with an overall $\Delta\chi^2 = -35.72$ compared to $\Lambda$CDM.
Furthermore, we detect variations in $\alpha$ and $m_{\rm e}$, yet the binding energy remains consistent with the Standard Model, affirming the precise timing of visible recombination and the insensitivity of observables to specific variations in fundamental parameters at first order.
Additionally, the FCV represents the phenomenological need to scale the Thomson scattering rate accordingly when exploring models with higher $H_0$, regardless of the mechanism which amplifies it.
The incorporation of dark radiation and ADM upholds the ratios of matter to radiation, pressure-supported to pressureless matter, and fluid to free-streaming radiation, in alignment with $\Lambda$CDM predictions.
The required ADM fraction is $f_{\rm adm} = 0.0322\pm0.0103$, and the mirror sector is found to be slightly cooler than the visible sector with $\xi_{\rm dark} = 0.6849\pm0.07$.
The variation in neutrino energy density, $N_{\rm ur}$, yields best-fit values about 10\% higher than the Standard Model predictions and nicely matches the predicted scaling of $N_{\rm ur} \rightarrow \lambda^2 N_{\rm ur}$.
However, across all likelihood groups, the results consistently align with $N_{\rm ur} = 3.044$ within the 68\% confidence level (CL).

Figure \ref{fig:spectra} demonstrates the CMB temperature-temperature (TT) and polarization-polarization (EE) spectra for the best-fit values obtained for the mirror model when considering likelihood group C.
We find that even with significantly enhanced values of $H_0 = 73.80$ km/s/Mpc compared to $\Lambda$CDM, we do not sacrifice agreement to CMB spectra observations both at low $\ell$ and high $\ell$ multipoles as seen in Table \ref{tab:chi2_comparison} and Fig.~\ref{fig:spectra}, although there is a slight trade for agreement for high $\ell$ from the low $\ell$ data.

\subsection{Profile Likelihood with \texttt{Procoli}}
\begin{figure}
\includegraphics[width=8.6cm]{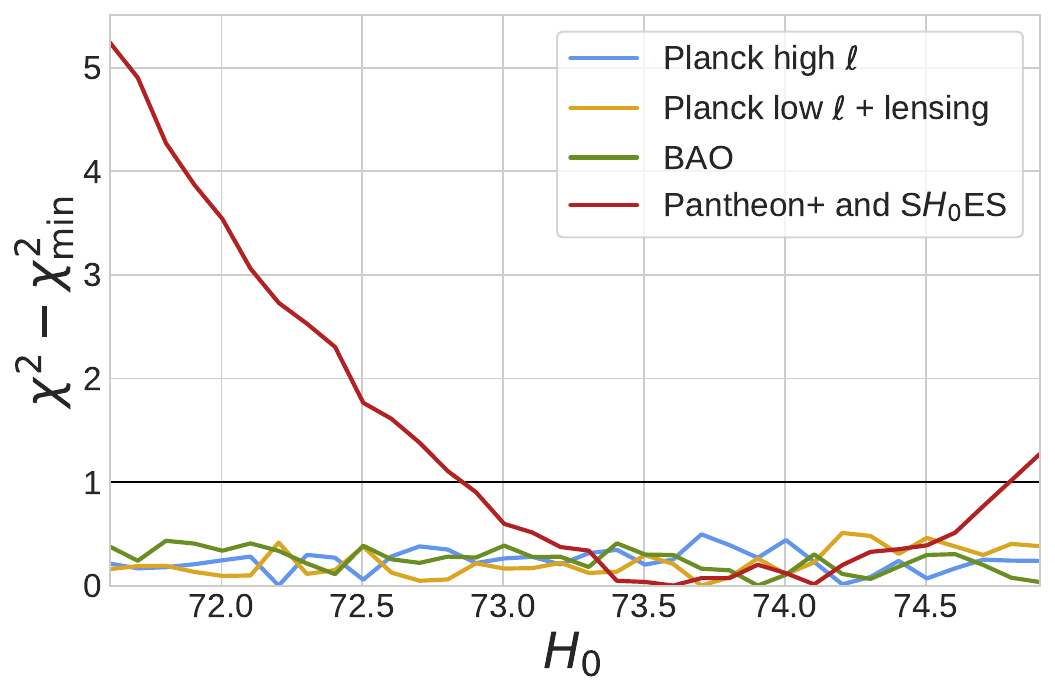}
\caption{This plot demonstrates the results from a profile likelihood scan over $H_0$ utilizing the \texttt{Procoli} package for the synchronous mirror recombination model when considering likelihood group C. The results are normalized to each datasets minimum $\chi^2$ value, allowing comparison between all data sets. We point out the $\Delta\chi^2 = 1$ line which represents the 68\% CL.}
\label{fig:pflkl}
\end{figure}
To delve deeper into the connection between preserving ratios and cosmological observables, we perform a profile likelihood scan over $H_0$ across all three likelihood groups utilizing the newly developed \texttt{Procoli} package \cite{Karwal:2024qpt}.
Figure \ref{fig:pflkl} displays the outcomes of one of these scans, for likelihood group C, demonstrating that CMB and LSS data are insensitive to changes in $H_0$ while exploring along the FFAT symmetry direction.
The degeneracy is only resolved when a calibrating measurement of $H_0$ is incorporated.
In table \ref{tab:chi2_comparison}, we present a comparison of the $\chi^2$ values across each dataset within likelihood group C for three distinct cases: $\Lambda$CDM, the synchronous mirror recombination model, and the asynchronous mirror recombination model (see Sec.\ref{sec:brokenmirror}).
We observe that both mirror models result in a reduced $\chi^2$ for the Planck high-$\ell$ measurements and for the Pantheon+ data, once calibrated by the S$H_0$ES measurement. 
The model's effectiveness at high $\ell$ stems from maintaining the $\dot{\kappa}/H$ ratio. 
This preservation allows for higher values of $H$, driven by the increased ADM and dark radiation density, ensuring alignment with the Pantheon+ with S$H_0$ES dataset.
The fit to the low $\ell$ TT spectra and the BOSS DR12 BAO dataset shows a slight deterioration. 
However, measurements from low redshift BAO largely compensate for this change.
\begin{table}
    \centering
    \begin{tabular}{cccc}
        \hline
                                     & $\Lambda$CDM & Synchronous  & Asynchronous \\ &&Recombination&Recombination \\
        \hline
        \hspace{2 mm}Planck High $\ell$\hspace{2 mm} & \hspace{1 mm}2347.79\hspace{1 mm} & \hspace{1 mm}2343.97 \hspace{1 mm}& \hspace{1 mm}2342.18\hspace{1 mm} \\
        Planck Low $\ell$ TT         & 22.63        & 24.16   & 23.89   \\
        Planck Low $\ell$ EE         & 396.95       & 396.09  & 396.05  \\
        Planck Lensing               & 8.85         & 8.62    & 8.70    \\
        BOSS DR12                    & 3.57         & 4.28    & 4.41    \\
        BOA low z                    & 2.32         & 1.23    & 1.198   \\
        Pantheon+S$H_0$ES                   & 1319.44      & 1287.49 & 1287.45 \\
        \hline
        \hline
        $\Delta\chi^2$               & 0            & -35.72  & -37.65   \\
        \hline
    \end{tabular}
    \caption{This table provides a comparison of the fit to individual datasets within likelihood group C across three models: the baseline $\Lambda$CDM, the synchronous mirror recombination model, and the asynchronous mirror recombination model. The $\Delta\chi^2$ value quantifies the change in $\chi^2$ relative to the baseline $\Lambda$CDM model.}
    \label{tab:chi2_comparison}
\end{table}

\subsection{\label{sec:coldthumbs}Cold Mirror Sector Volume Effects}

The one and two dimensional posteriors for likelihood groups A and B are influenced by a potentially large volume effect in the prior space in which $f_{\rm adm}$ becomes unconstrained when the mirror sector is very cold.
This result was first pointed out by ref.~\cite{Hughes:2023tcn} and is most clearly seen when considering likelihood group B.

\begin{figure}
\includegraphics[width=8.6cm]{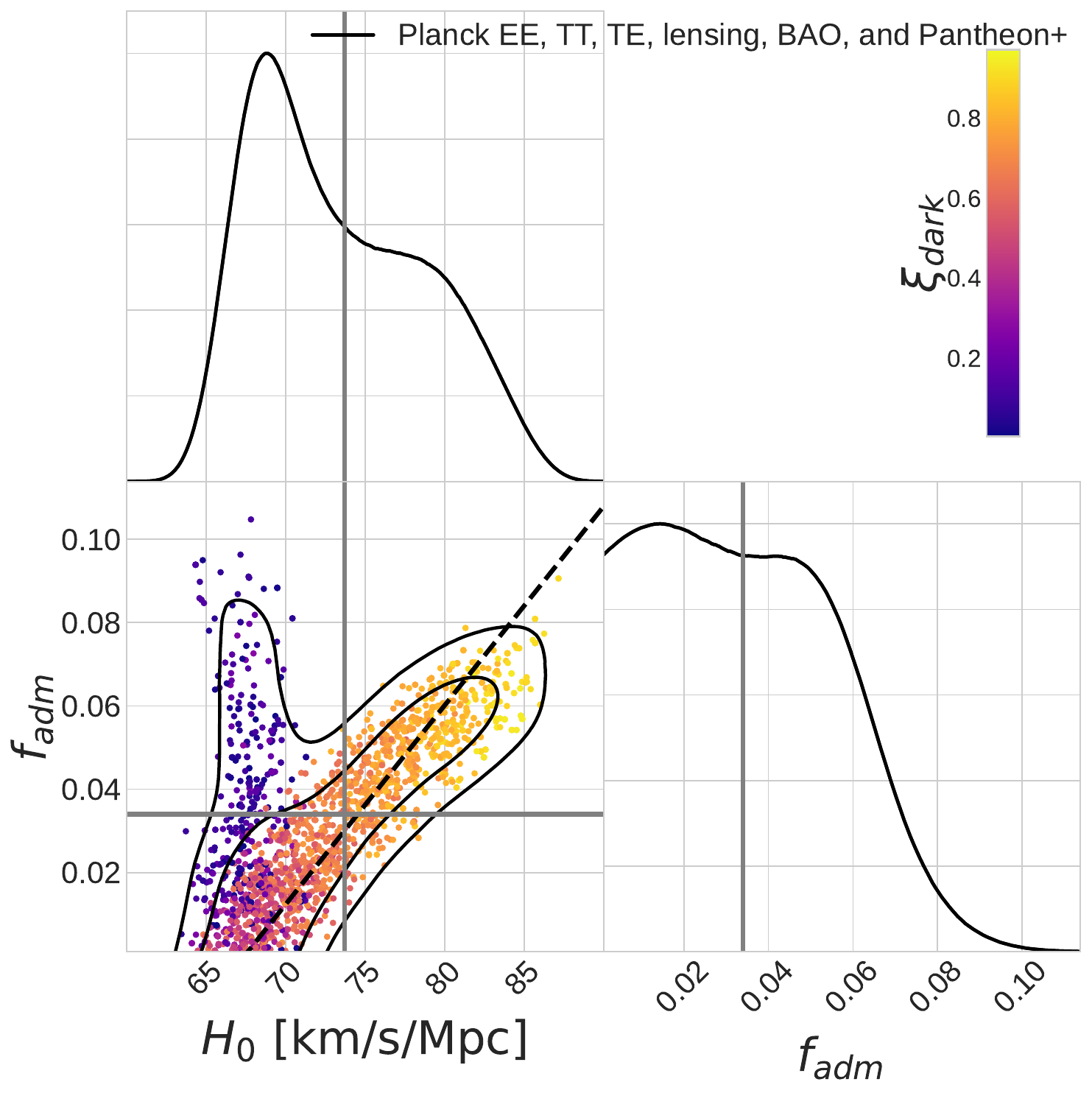}
\caption{This triangle plot showcases a volume effect shaping the posteriors for likelihood groups A and B in the synchronous mirror recombination model. The one-dimensional posteriors, prominently influenced by this volume effect, display peaks that are shaped accordingly. A black dashed line in the plot indicates the FFAT symmetry direction. The color scale indicates the temperature of the mirror sector, with purple colors being very cold mirror sectors. The grey lines indicate best fit values, and because of the volume effect introduced by cold dark sectors, do not necessarily align with the peak of the one dimensional distribution.}
\label{fig:coldds}
\end{figure}

In scenarios where the mirror sector temperature is cold ($\xi_{\rm d} \lesssim 0.25$), $f_{\rm adm}$ becomes unconstrained as the ADM begins to simply resemble the background cold dark matter. 
This similarity introduces a significant volume to the parameter space, echoing $\Lambda$CDM characteristics and resulting in `mitten' shaped two-dimensional contours with `cold thumbs'.
Figure \ref{fig:coldds} demonstrates that the synchronous mirror recombination model follows the expected FFAT symmetry direction for $f_{\rm adm}$ in the `mitten' when considering likelihood group B.
However, as $\xi_{\rm d} \rightarrow 0$, $f_{\rm adm} \rightarrow 1$ and the volume effect is introduced creating `cold thumbs' of the mitten.
This volume effect is removed when calibrating the Pantheon+ supernovae with S$H_0$ES, as the S$H_0$ES anchor strongly disfavours cold mirror sectors.
Ultimately, this increased volume influences the shape of the one dimensional posteriors when considering likelihoods groups A and B, but do not influence the best-fit parameters.
For instance, in figure \ref{fig:coldds}, we point out the best-fit values and note they do not align with the peaks of the distribution.

\subsection{\label{sec:brokenmirror}Additional Mirror Sector Freedom} 

As a final test of $\mathbb{Z}_2$ inspired ADM models, we now consider a scenario where the mirror sector's recombination is asynchronous with the visible sector in an asynchronous mirror recombination model. 
By varying the mirror binding energy to mirror sector temperature ratio through changes in the dark fine structure constant, we determine the recombination epoch for the mirror sector: ratios greater (less) than unity indicate earlier (later) mirror sector recombination compared to the visible sector.

\begin{figure*}
\includegraphics[width=\textwidth]{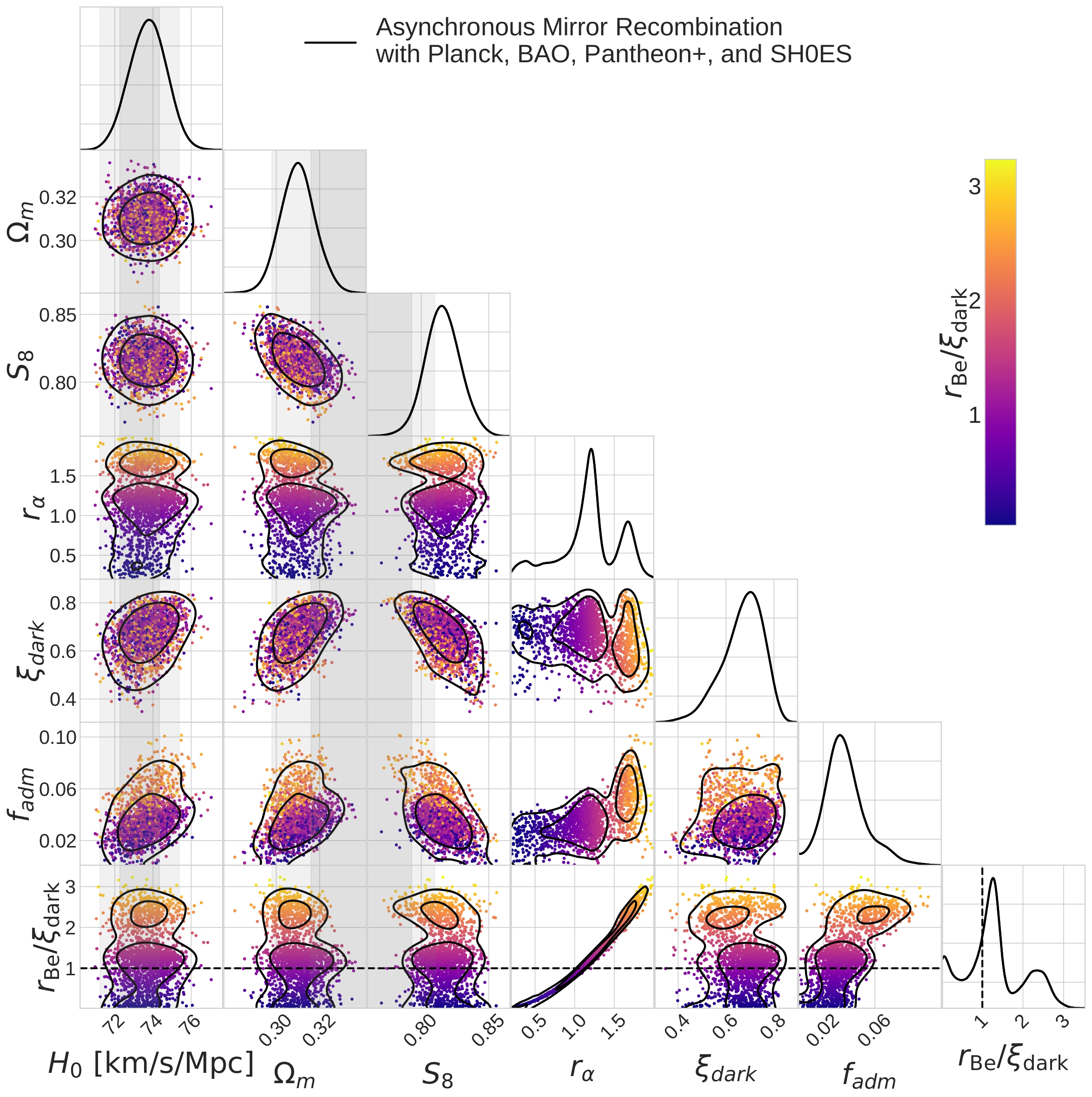}
\centering
\caption{This triangle plot showcases the asynchronous mirror recombination model. It displays three free mirror sector parameters alongside three key observables ($H_0$, $S_8$, and $\Omega_{\rm m}$) linked to cosmological tensions. 68\% (dark grey) and 95\% (light grey) CL limits for $H_0$, $S_8$, and $\Omega_{\rm m}$ appear in the relevant plots. The color coding reflects the ratio of the mirror sector binding energy to its temperature relative to visible sector values, a determinant of the mirror sector's recombination epoch. Ratios above one suggest earlier recombination compared to the visible sector, while values below one indicate a later recombination. The black dashed line indicates the values which align with the visible sector binding energy to temperature ratio. }
\label{fig:1dscontour}
\end{figure*}

Our findings reveal a statistical preference in $\Delta \chi^2$ for an earlier mirror sector recombination, finishing just before the onset of visible hydrogen recombination.
Recall that in this model, the mirror sector interacts with the visible sector only gravitationally.
Prior to mirror sector recombination, dark photons act as a fluid. 
Once the majority of atomic dark matter recombines, the dark photons start to freestream.
Therefore, CMB data seems to favor an injection of freestreaming radiation before hydrogen recombination but after helium recombination which would have the effect of gravitationally pulling on sound horizon wavefronts, and efficiently transporting energy out of the system \cite{Bashinsky:2003tk,Cyr-Racine:2013jua}. 
The asynchronous mirror recombination model with likelihood group C finds a $\Delta \chi^2 = -37.65$ compared to the $\Lambda$CDM model, and $H_0$ values which agree with local measurements.
Figure \ref{fig:1dscontour} illustrates the correlation between mirror sector parameters and cosmological tensions for the asymmetric recombination model, with parameter values given in table \ref{tab:brokenmirror}.
A bimodal feature related to the timing of mirror recombination is found, where it is disfavoured to recombine during visible HeII to HeI recombination.
As a general feature of this model and the synchronous mirror model, the ADM experiences pressure support from the dark radiation until the end of its drag epoch, slowing the infall of dark matter into initial seeds of structure.
This allows the one and two dimensional posteriors in figure \ref{fig:1dscontour} to penetrate into higher values of $\Omega_{\rm m}$ and lower values of $S_8$.
We note that this preference for later recombination in the dark sector is phenomenologically similar to recent studies showing that allowing the free electron fraction during visible recombination to vary freely suggests a non-standard recombination history, with part of the visible sector recombining later than predicted by the standard model \cite{Lynch:2024gmp, Lynch:2024hzh}.
Finally, while outside the scope of this study, it may be of great interest to investigate how subdominant components of the mirror sector adhering to similar ADM phenomenology influence the growth of structure on scales relevant for the $S_8$ tension.
Local observations of satellite galaxies and searches for thin dark matter disks constrain the energy dissipation rate caused by ADM additions to the dark sector and its impact on the formation and shape of these small nearby objects, providing valuable guidance for future dark sector model development \cite{Schutz:2017tfp,Gemmell:2023trd,Roy:2023zar,Lin:2024fyw,Rose:2024xcb}.
\begin{table}
    \centering
    \begin{tabular}{ c  c  c }
    \hline
        & Parameter & Asymmetric Recombination \\
    \hline
    $\Lambda$CDM& $h$& $\hspace{5 mm}0.7375\hspace{1 mm}(0.7381)\pm0.0100\hspace{5 mm}$ \\
    
        & $\omega_{\rm b}$    & $0.0224\hspace{1 mm}(0.0022)\pm0.0003$ \\
        
        & $\omega_{\rm cdm}$  & $0.1461\hspace{1 mm}(0.1474)\pm0.0067$ \\
        
        & $n_{\rm s}$         & $0.9624\hspace{1 mm}(0.9591)\pm0.0106$ \\
        
        & $A_{\rm s}10^9$     & $2.0889\hspace{1 mm}(2.0718)\pm0.0460$ \\
        
        & $\tau_{\rm reio}$   & $0.0519\hspace{1 mm}(0.0515)\pm0.0074$ \\
        
    \hline
    ADM    & $\xi_{\rm dark}$    & $0.6715\hspace{1 mm}(0.7132)\pm0.0817$ \\
    
           &$f_{\rm adm}$        & $0.0362\hspace{1 mm}(0.0347)\pm0.0143$ \\
           
           &$r_{\alpha}$ & $1.1804\hspace{1 mm}(1.2127)\pm0.3811$\\
           
    \hline
    FCV    &$\delta_{\alpha}$   & $0.0122\hspace{1 mm}(0.0142)\pm0.0044$ \\
    
        &$\delta_{\rm me}$   & $-0.0254\hspace{1 mm}(-0.0312)\pm0.0140$ \\
        
    \hline
    Other   & $N_{\rm ur}$    & $3.3757\hspace{1 mm}(3.2859)\pm0.3628$\\
    \hline
    \hline
Dark    &$r_{\rm me}$ & $0.5721\hspace{1 mm}(0.6132)\pm0.0950$\\
        
        &$r_{\rm mp}$ & $0.6715\hspace{1 mm}(0.7132)\pm0.0817$\\

        &$r_{\rm Be}/\xi_{\rm dark}$   & $1.3077\hspace{1 mm}(1.2645)\pm0.7095$\\
     \hline
 Visible&$\delta_{\rm Be}$   & $-0.0015\hspace{1 mm}(-0.0035)\pm0.0084$\\
        &$\delta_{\sigma_{T}}$ & $0.0796\hspace{1 mm}(0.0959)\pm0.0385$\\
        &$\Omega_{\rm m}$     & $0.3098\hspace{1 mm}(0.3117)\pm0.0827$\\
        
        &$S_8$           & $0.8161\hspace{1 mm}(0.8139)\pm0.0131$\\
     
        &$z_{\rm rec}$        & $1090.1\hspace{1 mm}(1088.1)\pm9.3337$\\

    \hline
    \hline
         &$\Delta \chi^2$    & $-37.65$ \\
    \hline
    \end{tabular}
    \caption{This table presents the mean (best-fit) values for the asynchronous mirror recombination model when considering likelihood group C, with all errors representing 68\% CL. The format is the same as in table \ref{tab:my_label}.}
    \label{tab:brokenmirror}
\end{table}

\begin{figure}
\includegraphics[width=8.6cm]{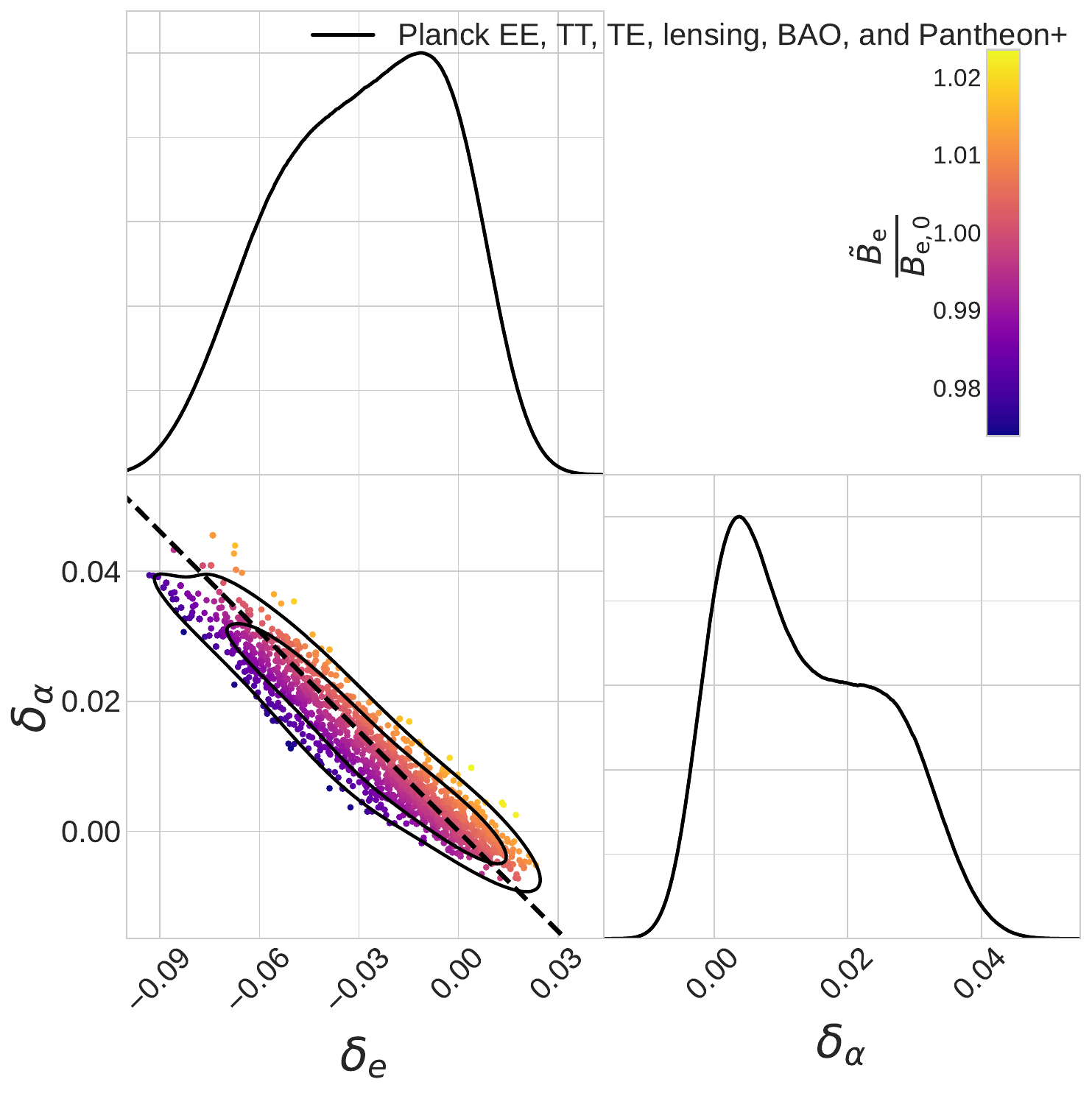}
\caption{This triangle plot demonstrates the relationship between variations in $\alpha$ and $m_{\rm e}$ in the synchronous mirror recombination model. A black dashed line marks the Standard Model binding energy. The samples are plotted as points with colors corresponding to the ratio of the varied visbile binding energy compared to the Standard Model value.}
\label{fig:bindingenergy}
\end{figure}

Finally, an interesting coincidence has appeared in the analysis of the synchronous and asynchronous mirror recombination models when considering likelihood groups B and C.
The best-fit temperature for the mirror sector in the asymmetric recombination model is $\xi_{\rm dark} = 0.7132$ or 1.944 K. 
This temperature is strikingly similar to the 1.95 K predicted for the neutrino background.
Note that this is not simply adding an additional neutrino species and naming it something else, as this radiation behaves as a fluid until mirror recombination.
This coincidence bears intriguing implications, connecting back to the second clue mentioned in Section \ref{sec:intro}.
This observation could imply that, at some point in cosmic history, the mirror sector radiation might have coupled to the neutrino or visible sector, allowing the two sectors to achieve thermal equilibrium with one another and explaining this temperature coincidence \cite{Ahmed:2023vdb, Giovanetti:2024orj}.
A similar phenomenology to this was realized in \cite{Berlin:2017ftj}.

\subsection{Constant Binding Energy}

A particular scaling of $\alpha$ and $m_{\rm e}$ has been determined to maintain constant binding energy while enabling increased Thomson scattering rates \cite{Greene:2023cro}.
Without prior knowledge of this specific strategy, the MCMC sampler naturally navigates this path, revealing the insensitivity of cosmic observables to individual changes in either $\alpha$ or $m_{\rm e}$, focusing instead on maintaining the invariance of the binding energy within the visible sector.
This degeneracy direction is depicted in Figure \ref{fig:bindingenergy}, where the plot's two-dimensional posterior closely follows lines of constant binding energy, and guarantees that recombination in the visible sector occurs at a time consistent with $\Lambda$CDM predictions, thereby preserving distance measurements.
Simultaneously, the Thomson scattering rate not left invariant in the direction of constant binding energy, allowing the sampler to explore regimes with constant $\frac{\dot{\kappa}}{H}$ ratios.
We show likelihood group B for its capacity to explore the FFAT symmetry's degeneracy direction without the S$H_0$ES calibration which anchors the result.

\section{\label{sec:conc} Conclusion}

In this study, we leverage insights into the dark sector's connection to the visible sector, and the predictions of ratios by $\Lambda$CDM.
We introduce a streamlined, universal model of ADM, inspired by recent works investigating mirror sectors. 
In this model, a subdominant component of the dark sector ($\sim 3\%$) mirrors the visible sector's $U(1)$ symmetry for electromagnetic interactions.
Notably, with parameters analogous to those in the visible sector, the ADM is expected to experience a recombination event within the mirror sector at epochs parallel to those observed in the visible sector.
To assess the influence of adding a dynamic, subdominant ADM component to the mirror sector, we employ the FFAT symmetry to guide our modeling approach. 
This strategy safeguards cosmological observables, guaranteeing alignment with CMB and LSS data.
ADM inherently fulfills one of the FFAT symmetry's phenomenological requirements by contributing proportional amounts of matter, radiation, and cold dark matter densities to the Universe.
The second criterion of the FFAT symmetry involves maintaining the ratio of the Thomson scattering rate to the background expansion rate.
To achieve this, we suggest a variation of fundamental constants in a way that scales the Thomson scattering rate appropriately, while keeping the binding energy of hydrogen and helium constant. 
We discover that the CMB observational agreement remains unaffected by particular variations in $\alpha$ and $m_{\rm e}$, provided the binding energy remains unchanged.
While integrating fundamental constant variation into a realistic Lagrangian presents challenges, the underlying phenomenology remains essential. 
A model capable of adjusting the Thomson scattering rate similarly, without relying on FCV, would offer a compelling route to explore the concordant cosmological models.
While this study demonstrates the potential of mirror sector models to address certain cosmological tensions, it is important to note that the consistency of these models with BBN is not fully settled. 
Refs.~\cite{Berlin:2017ftj,Berlin:2019pbq} provide a plausible route to address BBN consistency but requires the introduction of new couplings or mediators between the SM and dark sectors.
We hope that this work motivates more detailed studies examining the interplay of BBN and the implications for the viability of mirror sector models.

To investigate this phenomenology and its relationship to cosmic tensions, we conduct a MCMC analysis utilizing a modified version of \texttt{CLASS} which incorporates ADM.
We explore two distinct cases of mirror sector recombination: i) the synchronous mirror recombination model depicting the mirror sector recombining concurrently with the visible sector and ii) the asynchronous mirror recombination model allowing the mirror sector to undergo recombination independently from the visible sector.
Our findings indicate that under both recombination scenarios, the CMB and LSS observables show almost no sensitivity in response to changes in $H_0$.
Incorporating data from the local measurement of $H_0$, we determine a cosmologically inferred value of $H_0 = 73.8 \pm 1.0$ km/s/Mpc for both models.
This results in improved fits, especially for the high $\ell$ Planck data and Pantheon+ dataset, corresponding to a $\Delta\chi^2 = -35.72$ for the synchronous model and $\Delta\chi^2 = -37.65$ for the asynchronous model.
The asynchronous mirror recombination model reveals a preference for the mirror sector to undergo recombination slightly earlier than the visible sector. 
This timing places mirror sector recombination between the conclusion of helium recombination and the onset of hydrogen recombination.

This dual approach to investigating the Hubble tension has achieved fruitful results, summarized as follows:
\begin{itemize}
    \item The inclusion of a subdominant (3\%) atomic dark matter component in the mirror sector with a temperature of $\xi_{\rm d} \sim 0.7$, inspired by $\mathbb{Z}_2$ symmetries, alongside a mechanism that maintains a consistent $\dot{\kappa}/H$,  eliminates the tension in $H_0$.
    \item The inclusion of extra mirror sector species, with mass and coupling strengths mirroring those in the visible sector, results in mirror sector recombination occurring at times similar to that of the visible sector.
    Given the freedom to recombine independently, the mirror sector exhibits a preference for undergoing recombination just before the visible sector hydrogen recombines.
    \item By scaling the Thomson scattering rate in accordance with the background expansion rate, we achieve cosmological concordance. Our approach employs a phenomenological variation of fundamental constants, demonstrating that targeted adjustments to $\alpha$ and $m_{\rm e}$ facilitate the necessary scaling while maintaining compatibility with CMB observations
\end{itemize}
Our work has highlighted several intriguing potential links between mirror models and cosmological tensions while offering avenues for further investigation into possible dark sector interactions.
Significant efforts are required from both particle physics and cosmology disciplines to further verify the efficacy of this or similar models at addressing cosmological tensions.
However, the phenomenology is clear: $3\%-4\%$ of the dark sector behaving as ADM while ensuring the ratio of $\dot{\kappa}/H$ is left invariant compared to $\Lambda$CDM produces concordant cosmology. 

\acknowledgments
We would like to thank David Camarena for his support and useful comments on the draft, as well as Ryan Janish for useful discussions.
This work was supported by the National Science Foundation (NSF) under grant AST-2008696. F.-Y. C.-R. would like to thank the Robert E.~Young Origins of the Universe Chair fund for its generous support. 
We would like to thank the UNM Center for Advanced Research Computing, supported in part by the NSF, for providing the research computing resources used in this work. 
This material is based upon work supported by the U.S. Department of Energy, Office of Science, Office of Workforce Development for Teachers and Scientists, Office of Science Graduate Student Research (SCGSR) program. The SCGSR program is administered by the Oak Ridge Institute for Science and Education for the DOE under contract number DE‐SC0014664.

\appendix

\mycomment{
\section{Affleck-Dine Condensates}
In this study, we deliberately maintain a broad parameter space to concentrate on exploring the phenomenology and to not introduce model bias.
However, it's important to highlight the connection between our work and the findings presented in reference \cite{Blinov:2021mdk}.
In reference \cite{Blinov:2021mdk}, the authors examine a phenomenologically similar mirror-like model populated through a mechanism of Affleck-Dine condensation.
This model offers a natural explanation for the precise adjustments required for the dark sector to recombine approximately concurrently with the visible sector, grounded in the correlations between the vacuum expectation values of both sectors.
An Affleck-Dine condensate mechanism for baryogenesis in the dark sector can satisfy several criteria: it results in lighter fundamental particles in the dark sector, maintains similar binding energy to temperature ratios between both sectors, and ensures that the matter-radiation ratio in the dark sector mirrors that in the visible sector.
These criteria align with the requirements previously outlined! 
Utilizing this mechanism to populate the dark sector naturally fulfills the desired phenomenology.
Models like the one presented in Refs.~\cite{Blinov:2021mdk,Berlin:2019pbq} can avoid thermalization of the dark sector and can circumvent the constraints imposed by $\Delta N_{\rm eff}$ from Big Bang Nucleosynthesis (BBN). 
However, as this is beyond the scope of our investigation, we merely mention the connection and currently set aside this issue and mandate that the helium abundance align with BBN predictions.
}

\section{Constraints on FCV}\label{ap:FCVconstraints}
Adjusting fundamental constants introduces challenges that span both theoretical and observational domains.
Although we have pinpointed a variation in the CMB during recombination that evades detection, geological records, atomic spectra, quasar observations, and potentially BBN impose much tighter constraints.
To ensure the model is not in direct disagreement with current observables, we briefly review the main methods for constraining FCV.
For an extensive review, see refs.~\cite{Uzan:2010pm, Martins:2017yxk}. 

Geological constraints, assessing periods comparable to Earth's age (or $z < 0.5$), are based on measuring relative abundances of various heavy element isotopes. 
The most stringent of these arises from the fascinating Oklo natural reactor phenomenon, where natural nuclear fission occurred approximately 1.7 billion years ago \cite{Shlyakhter:1982yr}. 
Analysis of this event, considering the neutron mass and decay resonance energy, has significantly constrained the variation of $\alpha$ to $|\Delta \alpha/\alpha| < 1.1 \times 10^{-8}$ over this period \cite{Davis:2015ila}. 

Atomic clocks provide a highly precise method for measuring FCV today ($z=0$).
These experiments seek to detect variation in frequencies of long term, stable oscillators and probe the nature of $\alpha$, the electron to proton mass ratio $\mu$, and the proton gyromagnetic ratio $g_{\rm p}$.
By measuring frequencies related to the gross, fine-structure, and hyper-fine structure atomic transitions they measure combinations of $R_{\infty}$, $\alpha^2 R_{\infty}$, and $\mu\alpha^2R_{\infty}$, respectively, where $R_{\infty}$ is the Rydberg constant.
The scaling of $\alpha$ and $m_{\rm e}$ proposed in this study specifically leaves the Rydberg constant and $\mu\alpha^2$ invariant, which could leave some atomic clock measurements insensitive to this specific variation.
However, the fine structure measurements ($\alpha R_{\infty}$) effectively break the degeneracy.
Current atomic clock data restricts the variability of $\alpha$ to about $|\dot{\alpha}/\alpha| < 10^{-19}$ annually \cite{Filzinger:2023zrs}. 

Quasar (QSO) absorption spectra provide the earliest directly observable constraints on FCV. 
As some of the Universe's most luminous sources, QSOs allow observations up to redshifts of $z \sim 7$, reaching back to less than 1 Gyr after the Big Bang. 
Astronomers analyze absorption spectra from cold gas clouds along QSOs' line of sight up to $z \sim 3$ (about 2 Gyr old), setting constraints on early Universe FCV. 
This method, however, introduces additional complexity due to the Universe's expansion redshifting the spectra. 
FCV constraints using QSOs rely on identifying chromatic effects in the spectra, impacting all wavelengths and may be difficult to disentangle from redshift.
Current constraints from QSO absorption spectra separately limit $\alpha$ variation to $|\Delta \alpha/\alpha| < 10^{-6}$ at $z \sim 2.4$ \cite{Murphy:2016yqp} and $|\Delta \mu/\mu| < 10^{-5}$ at $z \sim 0.9$ \cite{Kanekar:2014ota}.

Big Bang nucleosynthesis offers potential constraints on FCV mere minutes after the end of inflation at redshifts of $z \sim 10^8$. 
Standard BBN models, which successfully predict light elemental abundances, potentially anchor FCV constraints to this early epoch. 
BBN's sensitivity extends to various factors, including the expansion rate, neutron-proton number density ratio and mass difference, neutron lifetime, and the Coulomb barrier strength. 
This complexity opens a challenging avenue, as BBN's outcomes are not clear when considering variations of $\alpha$ and $m_{\rm e}$.
However, a comprehensive analysis of FCV's impact on BBN demands not only advanced nuclear reaction rate calculations but also detailed lattice QCD and QED computations \cite{Coc:2006sx,Coc:2012xk,BMW:2014pzb, Clara:2020efx}. 

We summarize the constraints on FCV as follows:
\begin{itemize}
\item Atomic clocks confirm the stability of $\alpha$ and $m_{\rm e}$ at present ($z = 0$) and that linear variations are ruled out.
\item Geological and quasar (QSO) observations suggest no significant variation in these parameters individually over the past 10 billion years ($z \sim 3$).
\item The exact impact of the FFAT symmetry combined with FCV on Big Bang Nucleosynthesis outcomes remains uncertain at $z \sim 10^8$. See App.~\ref{app:BBN} for more discussion.
\end{itemize}
However, it's interesting to highlight that most discussions herein typically focus on varying a single parameter. 
Simultaneously varying both $\alpha$ and $m_{\rm e}$ might weaken some constraints, particularly if they primarily depend on the binding energy, which remains unchanged under this specific FCV. 
Under the most conservative assumption that simultaneous variations in $\alpha$ and $m_{\rm e}$ do not impact constraining power, a particle physics model aiming to realize the required Thomson scattering rate increase must vary these parameters while satisfying several criteria: it must preserve BBN predictions, avoid rapid variation compared to the timescale set by recombination, and align with Standard Model values before the earliest QSO observations, maintaining stability thereafter.

\section{\label{app:BBN}Comments on BBN and FCV}
Big Bang Nucleosynthesis is significantly influenced by two critical quantities that depend on the fine structure constant and the electron mass: the neutron to proton mass difference ($Q_{\rm np}$) and the neutron lifetime ($\tau_{\rm n}$). 
Ref.~\cite{Muller:2004gu} explores the impact of varying $\alpha$ and $m_{\rm e}$ linearly, establishing two key relationships:
\begin{equation}
\frac{\Delta Q_{\rm np}}{Q_{\rm np}} = -0.59 \frac{\Delta \alpha}{\alpha},
\end{equation}
\begin{equation}
\frac{\Delta \tau_{\rm n}}{\tau_{\rm n}} = 3.86 \frac{\Delta \alpha}{\alpha} + 1.52 \frac{\Delta m_{\rm e}}{m_{\rm e}}.
\end{equation}
In this analysis, we omit variations in the vacuum expectation value and quark masses as examined in \cite{Muller:2004gu}.
Applying the best-fit values identified in Section \ref{sec:results}, we ascertain that following the FFAT symmetry direction results in a decrease in $Q_{\rm np}$ by approximately 1\% and an increase in $\tau_{\rm n}$ by around 1\%.

BBN unfolds in three crucial stages, with the neutron to proton mass difference ($Q_{\rm np}$) and the neutron lifetime ($\tau_{\rm n}$) playing pivotal roles:
\begin{enumerate}
    \item Initially, the temperature remains high enough for the weak interaction rate to maintain protons and neutrons in equilibrium.
    \item Once the weak interaction rate dips below the Hubble rate, inverse reactions cease, and the neutron to proton ratio becomes fixed at an amount sensitive to $Q_{\rm np}$ and the background temperature. However, neutrons are unstable and will continue to decay according to $\tau_{\rm n}$ until the onset of light element synthesis.
    \item Finally, the temperature falls to a level where primordial nucleosynthesis begins, facilitated by the formation of deuterons.
\end{enumerate}
This simplified overview demonstrates that BBN is influenced by $Q_{\rm np}$ and $\tau_{\rm n}$, as well as the background expansion rate.
Without comprehensive calculations, pinpointing the exact impact of these factors on the predicted helium-4 yields is challenging and very model dependent.
However, it seems that an increase in $Q_{\rm np}$ and a decrease in $\tau_{\rm n}$ would increase the helium-4 yields.
Ultimately, BBN could serve as the decisive assessment for any proposed resolution to the Hubble tension, as elemental abundances offer stringent constraints on the expansion rate during that epoch.
For recent discussion regarding the connection between BBN and FCV see Refs.~\cite{Deal:2021kjs,Martins:2020syb,Clara:2020efx, Meissner:2023voo}.

\bibliography{apssamp}

\end{document}